\documentclass{aa}

\usepackage{graphicx}
\usepackage{natbib}
\usepackage{xcolor}

\bibpunct{(}{)}{;}{a}{}{,} 
\usepackage{txfonts}
\begin{document}

   \title{On the Effect of Rotation on Populations of
   Classical Cepheids\\
   I.~Predictions at Solar Metallicity}

   \subtitle{}

   \author{Richard I. Anderson$^1$
          \and
          Sylvia Ekstr\"om$^1$
          \and
          Cyril Georgy$^2$
          \and
          Georges Meynet$^1$
          \and
          Nami Mowlavi$^1$
          \and
          Laurent Eyer$^1$
          }
   \authorrunning{R.~I. Anderson et al.}

   \institute{$^1$Observatoire de Gen\`eve, Universit\'e de Gen\`eve, 51 Chemin des
   Maillettes, CH-1290 Sauverny, Switzerland\\
   $^2$Astrophysics group, Lennard-Jones Laboratories, EPSAM, Keele University,
   Staffordshire, ST5 5BG, UK\\
              \email{richard.anderson@unige.ch}
             }

   \date{Received 5 November 2013; accepted 6 February 2014}

 
  \abstract
   {Classical Cepheids are among the most important variable star types due to
   their nature as standard candles and have a long history of modeling in terms
   of stellar evolution. The effects of rotation on Cepheids have not
   yet been discussed in detail in the literature, although some qualitative
   trends have already been mentioned.}
   {We aim to improve the understanding of Cepheids from an evolutionary
   perspective and establish the role of rotation in the Cepheid paradigm. In
   particular, we are interested in the contribution of rotation to the problem
   of Cepheid masses, and explore \emph{testable} predictions of quantities that
   can be confronted with observations.}
   {Recently developed evolutionary models including a homogeneous and
   self-consistent treatment of axial rotation are studied in detail during the
   crossings of the classical instability strip (IS). The dependence of a suite
   of parameters on initial rotation is studied. These parameters include mass,
   luminosity, temperature, lifetimes, equatorial velocity, surface abundances,
   and rates of period change.}
   {Several key results are obtained:
  i)\,mass-luminosity (M-L) relations depend on rotation, particularly during
  the blue loop phase; ii)\,luminosity increases between crossings of
  the IS. Hence, Cepheid M-L relations at fixed initial rotation rate depend on
  crossing number (the faster the rotation, the larger the luminosity difference
  between crossings); iii)\,\emph{the Cepheid mass discrepancy
  problem vanishes} when rotation and crossing number are taken into account,
  without a need for high core overshooting values or enhanced mass loss;
  iv)\,rotation creates dispersion around average parameters predicted at fixed
  mass and metallicity. This is of particular importance for the
  period-luminosity-relation, for which rotation is a source of intrinsic
  dispersion; v)\,enhanced surface abundances do not unambiguously distinguish
  Cepheids occupying the Hertzsprung gap from ones on blue loops (after
  dredge-up), since rotational mixing can lead to significantly enhanced Main
  Sequence (MS) abundances; vi)\,rotating models predict greater Cepheid ages
  than non-rotating models due to longer MS lifetimes.
   }
   {Rotation has a significant evolutionary impact on classical Cepheids and
   should no longer be neglected in their study.}


   \maketitle
%
\section{Introduction}
Classical Cepheids are objects of interest for many areas of
astrophysics. On the one hand, they are excellent standard candles allowing the
determination of distances in the Milky Way and up to Virgo cluster distances. On
the other hand, they are excellent objects for constraining stellar evolutionary
models. Accordingly, Cepheids have played a special role among pulsating
(variable) stars and have a long history of modeling efforts both in terms of
their evolution and pulsations \citep[see][and references
therein]{2013IAUS..289..116B}.

Following the development of pulsation models in the late
1960's \citep{1969MNRAS.144..461S,1969MNRAS.144..485S,1969MNRAS.144..511S},
systematic differences between pulsational and evolutionary masses of Cepheids
became apparent. These differences were originally referred to as
Cepheid mass anomalies \citep{1980ARA&A..18...15C}. Improved opacities
\citep{1991ApJ...371..408I,1994MNRAS.266..805S} could mitigate a good fraction
of the disaccord in terms of masses. Nowadays, it is common to speak of  
\emph{the mass discrepancy} as the systematic offset between evolutionary masses
and those derived with other methods
\citep[e.g.][]{2006MmSAI..77..207B,2008ApJ...677..483K}, with a typical
disagreement at the level of $10-20\%$. 

Several mechanisms have been put forward to resolve the mass discrepancy
problem. Among the most prominent in the recent literature are augmented
convective core overshooting \citep{2012ApJ...749..108P} and pulsation-enhanced
mass-loss \citep{2008ApJ...684..569N}. However, the effect of rotation on
evolutionary Cepheid masses has not been discussed in detail although the impact
of rotation on the mass-luminosity and period-luminosity relations have already
been mentioned previously in the literature
\citep[see][]{2000ARA&A..38..143M,2001A&A...373..555M}.

The first large grid of models incorporating a homogeneous and consistent
treatment of axial rotation was recently presented by \citet[from
hereon:
paper\,I]{paperI}. Subsequently, \citet[from
hereon: paper\,II]{paperII} extended the grid in terms of
rotation for stars between $1.7$ and $15\,M_\odot$. Using these state-of-the-art grids it
is now possible, for the first time, to consider the effect of rotation on
populations of classical Cepheids. 

Cepheid progenitors are B-type stars on the Main Sequence (MS).
Observationally, B-type stars are known for their fast rotation since the first
homogeneous study of rotational velocities by \citet{1949ApJ...110..498S}.
Thus, it is empirically known that fast rotation is common for the progenitors
of Cepheids.
\citet{2010ApJ...722..605H} have recently carried out a very detailed
investigation of rotational velocities for B-stars of different masses and
evolutionary states (on the MS), providing even an empirical distribution of
rotation rates. Their distribution can serve as a guideline for our study in the
sense that the typical MS rotation rates of B-stars in the mass range
appropriate for Cepheids ($\sim 5 - 9 \,M_\odot$) is approximately
$v/v_{\rm{crit}}=0.3 - 0.4$, where $v_{\rm{crit}}$ denotes critical rotation
velocity. As shown in papers\,I\,\&\,II, such fast rotation can significantly
alter the evolutionary path of stars by introducing additional mixing effects
that impact MS lifetime, stellar core size, age, and MS surface abundances.
Clearly, these effects will propagate into the advanced stages of evolution,
such as the Cepheid stage.

The blue loop phase of intermediate-mass evolved stars during core helium
burning is very sensitive to the input physics
\citep[see the ``magnifying glass'' metaphor in][p. 305]{1994sse..book.....K}.
Thus, there is a two-fold interest in studying the models presented in papers\,I
\&\,II in terms of their predictions for Cepheids: i) properties of Cepheids
inferred using models (e.g. mass) can be updated to account for rotation; ii)
certain predictions made by the models can be tested immediately using
observational data (e.g. surface abundances).

This paper is the first in a series devoted to the endeavor of extending the
evolutionary paradigm of Cepheids to include the effects of rotation. We here
focus mainly on the detailed exploration of predictions made by the models as
well as their interpretation. A detailed comparison of these
predictions to observed features is in progress and will be presented in a
future publication for the sake of brevity. Further projects will include the
investigation of the combined effect of metallicity and rotation as well as a
self-consistent determination of pulsation periods and instability strip
boundaries.

The structure of this paper is as follows. In Sec.\,\ref{sec:Physics}, we briefly
state a few key aspects of the models presented in papers\,I\,\&\,II relevant
in the context of this work. Sec.\,\ref{sec:Predictions} contains the predictions
made by the rotating Cepheid models, focusing on the mass discrepancy
and features that are accessible to observations. Sec.\,\ref{sec:Discussion}
discusses the reliability and implications of the predictions made, and Sec.
\ref{sec:Conclusions} summarizes the key aspects.

\section{Description and comparison of input physics}
\label{sec:Physics}
The input physics of the stellar models are explained in detail
in paper\,I\,\&\,II. We refer the reader to these publications for the detailed
description of the models and summarize here only the most relevant features. 
The Solar composition adopted is that by
\citet{2005ASPC..336...25A}, and the detailed abundances of all
elements are listed in paper\,I.

Rotation is treated in the Roche model framework with the shellular hypothesis
as presented in \citet{1992A&A...265..115Z} and \citet{1998A&A...334.1000M}.
The shear diffusion coefficient is adopted from
\citet{1997A&A...321..134M} and the horizontal turbulence
coefficient is from \citet{1992A&A...265..115Z}.
Both were calibrated (see paper\,I) so that models presenting an averaged equatorial velocity during the
MS well in the observational range reproduce the mean surface enrichment of main sequence B-type 
stars at solar metallicity. Both were calibrated (see paper\,I) in order to
reproduce the mean surface enrichment of main sequence B-type stars at solar metallicity.

At the border of the convective core, we apply an overshoot parameter
$d_{\rm{over}}/H_P = 0.10$. This value was calibrated in the mass domain $1.35 -
9\,M_\odot$ at solar metallicity to ensure that the rotating models closely
reproduce the observed width of the MS band. Overshooting at the base of
the convective envelope is not explicitly included, although a particular choice
of the mixing length parameter\footnote{Here, we employ $l/H_P = 1.6$,
which reproduces the positions of both red giant branch and the red supergiant
stars, see Fig.\,2 in paper\,I.} in the outer convective envelope may
indirectly create a similar effect.

During the MS, intermediate-mass models ($M \leq 7\,M_\odot$) are evolved
without mass loss, $9$ to $12\,M_\odot$ models are evolved with the
\citet{1988A&AS...72..259D} mass-loss rates, and the more massive models with
the mass-loss recipe from \citet{2001A&A...369..574V}.
The radiative mass-loss rates used in the RSG and Cepheids phases are from
\citet{1975MSRSL...8..369R,1977A&A....61..217R} with a factor $\eta=0.5$ for
$M_{\rm{ini}} \leq 5.0\,M_\odot$ and $\eta=0.6$ for $M_{\rm{ini}}=7.0\,M_\odot$.
For $M_{\rm{ini}} \geq 9.0\,M_\odot$, we use the recipe from
\citet{1988A&AS...72..259D}, and for the RSG with $T_\mathrm{eff}<3.7$, we
use a fit presented in \citet{2001ASSL..264..215C}.
No pulsation-enhanced mass-loss is included in these static models.

In the following initial rotation velocities $v$ are stated relative to the 
critical velocity $v_{\rm{crit}}$, see paper\,I. Alternatively, initial rotation
rates are defined as $\omega = \Omega / \Omega_{\rm{crit}}$, see paper\,II. 
For reference, the average initial rotation speed of most B-stars is
$v/v_{\rm{crit}} = 0.4$ \citep{2010ApJ...722..605H}, which is equivalent to
$\omega = 0.568$. 

\subsection{Geneva and other Cepheid models}
The input physics and numerical implementation differ significantly between
different groups developing evolutionary models. To benchmark and provide
additional context for the Geneva models used here (paper\,I\,\&\,II), we
compare our evolutionary tracks with other references from the literature.    

\begin{figure}
\includegraphics[scale=0.37]{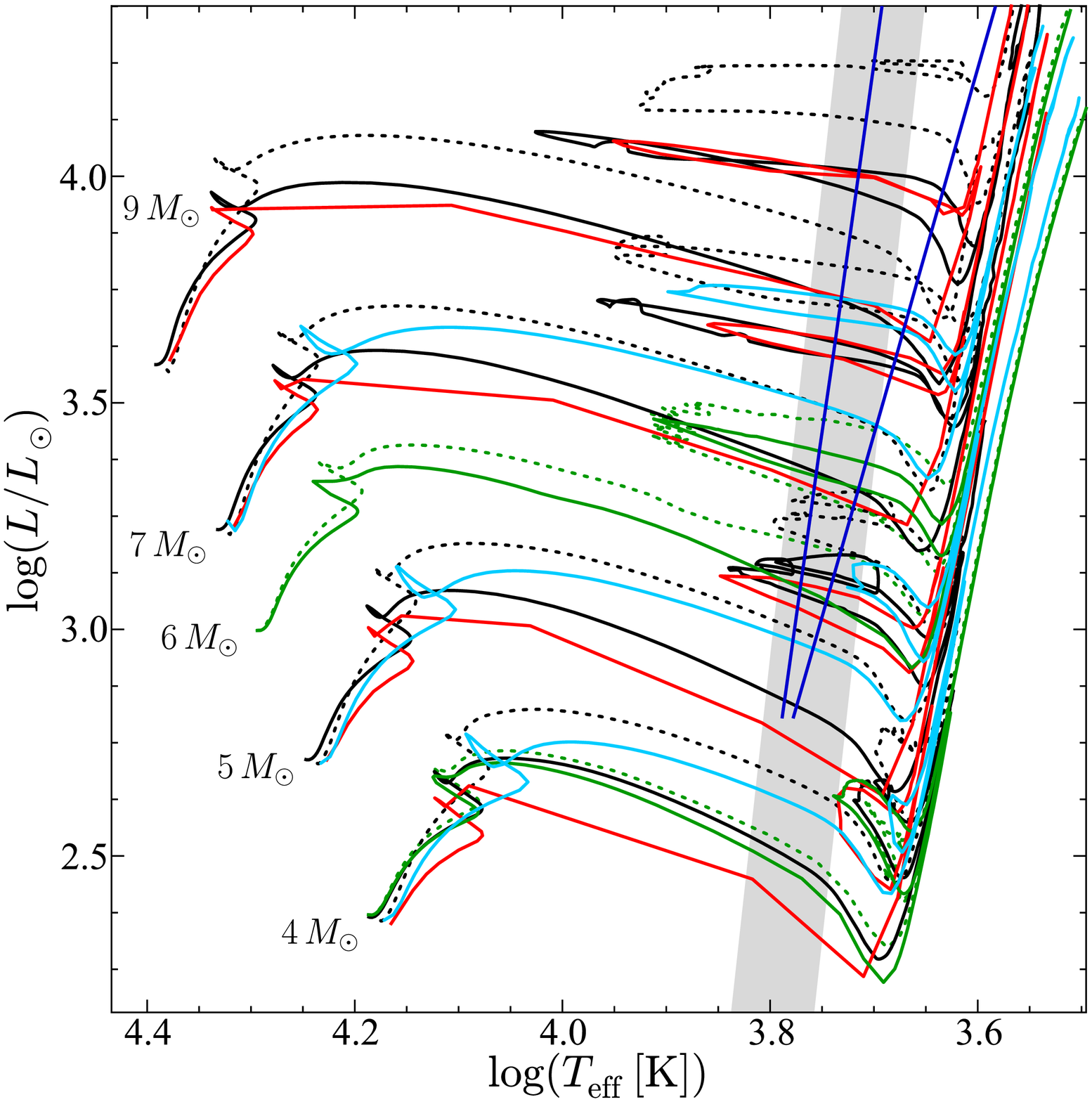}
\caption{Comparison of Geneva stellar evolutionary tracks 
(black solid lines : non-rotating; black dotted lines :
$v/v_{\mathrm{crit}}=0.4$) with Padova tracks
\citep[][cyan solid lines]{2008A&A...484..815B}, and tracks by  
\citet[][red solid lines]{1992ApJS...78..517C}
and \citet[green solid lines = non-rotating; green dotted lines =
$v/v_{\mathrm{crit}}=0.3$]{2012A&A...543A.108L}. The instability
strips by
\citet[wedge-shaped dark blue solid lines]{2000ApJ...529..293B} and 
\citet[gray shaded area]{2003A&A...404..423T} are also included.}
\label{fig:litModelComparison}
\end{figure}

Figure\,\ref{fig:litModelComparison} shows this comparison of non-rotating
(black solid lines) and rotating ($\nu/\nu_\mathrm{crit}=0.4$,
black dotted lines) Geneva tracks with Padova tracks \citep[][cyan
lines]{2008A&A...484..815B} and those used by \citet{2000ApJ...529..293B} in
their computation of instability strip boundaries used in the following
\citep[][red lines]{1992ApJS...78..517C}. Also shown are the non
rotating (green solid lines) and rotating
($\nu/\nu_\mathrm{crit}=0.3$, green dotted lines) models from
\citet{2012A&A...543A.108L}.    
 
Generally speaking, all models in Fig.\,\ref{fig:litModelComparison} have a lot
in common and differ mainly in the details. For the non rotating models,
the main sources of difference are the initial chemical composition (and thus
the opacity), and the value adopted for the overshooting parameter.  The
difference in composition and opacity slightly shifts the tracks on the ZAMS,
but in later phases, the main difference arises from the overshoot parameter.
The tracks from \citet{2008A&A...484..815B} show the high luminosity deriving
from an overshoot parameter equivalent to $2.5$ times the Geneva one, while the
tracks from \citet{1992ApJS...78..517C} do not have any overshoot during the
main sequence. 

The comparison of different evolutionary models in Fig.
\ref{fig:litModelComparison} clearly shows that the extent of the
blue loops is sensitive to the value used for the overshoot parameter: the
larger the overshooting parameter, the higher luminosity and the shorter the
loop. The rotating Geneva tracks predict consistently higher luminosity than any
of the other models in the figure, and should yield a lower mass limit for
Cepheids than the Padova tracks by \citet{2008A&A...484..815B}, judging from a
by-eye-interpolation.

\section{Properties of rotating stellar evolution models in the Cepheid stage}
\label{sec:Predictions}
In the following, the Cepheid stage refers simply to the portions
of the evolutionary tracks that fall inside the classical instability strip
(IS). %
A given evolutionary track can cross the IS three times, and thus may be
considered as a Cepheid during three different crossing numbers.
The first crossing occurs when the star evolves along the
Hertzsprung gap towards the Red Giant phase during a core contraction phase.
This crossing is expected to be very fast, and such Cepheids 
expected to be rare. Yet, some candidates have been reported in the
literature \citep[e.g.][]{2009AIPC.1170...59T}.
The majority of Cepheids are expected to be on the second and third
crossings. These stars are in the core helium burning phase and make up the
majority of a Cepheid's lifetime and are therefore the most likely to be
observed, see Sec.\,\ref{sec:lifetimes}.
We therefore focus our discussion mainly on Cepheids on second and third
crossings, neglecting first crossing Cepheids for the sake of brevity when it is
not essential.

Instability Strip (IS) boundaries have to be adopted in order to investigate the
predictions made by the models specifically for Cepheids. Several IS boundary
determinations can be found in the literature. We here adopt two such
definitions, one theoretical \citep{2000ApJ...529..293B}, and one
observational \citep{2003A&A...404..423T}. We selected these references because
they both provide analytical IS edge definitions for the range of metallicities
covered by the models.  

The IS boundaries by \citet{2000ApJ...529..293B} are based on 
limiting amplitude, non-linear, convective pulsation models that use 
evolutionary models by \citet{1992ApJS...78..517C} to predict luminosities for a
given mass.
We note that there are relevant differences between the models presented here
and those underlying the pulsation analysis by \citet{2000ApJ...529..293B}, notably
in the solar chemical composition, core overshooting, and, of course, the
fact that our models include rotation. It must therefore be kept in mind that
the present analysis uses IS boundaries that are not necessarily consistent
with the present models. A self-consistent determination of IS
boundaries and pulsation periods is foreseen for the near future.
Obviously, the choice of IS boundaries affects the range of
values predicted in the Cepheid stage such as  
Cepheid lifetimes, or equatorial velocities. However, other parameters such  
as luminosity should not be very sensitive to it. 

The typical rotation
velocity of Cepheid progenitors is approximately $v/v_{\rm{crit}}=0.3-0.4$ 
\citep{2010ApJ...722..605H}, which corresponds to a typical $\omega=0.5$. 
We therefore occasionally refer to $\omega=0.5$ as the `average'
rotation rate.
The initial hydrogen abundance, $Z_{\rm{ini}}$, is solar ($Z_\odot = 0.014$)
throughout this paper; a discussion of the combined effect of rotation and metallicity will be
presented in a future publication.

\subsection{Rotating Cepheids in the Hertzsprung-Russell Diagram}\label{sec:HRD}
\begin{figure}
\centering
\includegraphics{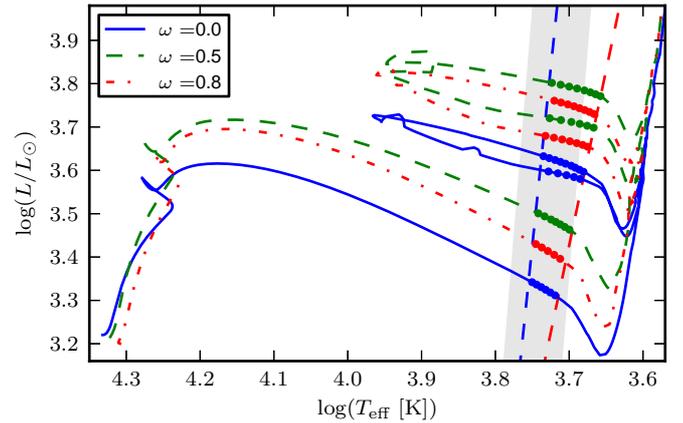}
\caption{Evolutionary tracks for $7\,M_\odot$ solar-metallicity models with
different initial rotation rates: $\omega = 0.0, 0.5, 0.8$ are drawn
as blue solid, green dashed and red dash-dotted lines, respectively. The red and
blue edge of the instability strip (IS) according to \citet{2000ApJ...529..293B}
is indicated by dashed lines. Grid points inside the IS are marked as solid
circles. The \citet{2003A&A...404..423T} IS is shown as a shaded
gray area.}
\label{fig:tracks}
\end{figure}

Figure\,\ref{fig:tracks} shows evolutionary tracks for $7\,M_\odot$ Solar
metallicity models at three rotation rates $\omega = 0.0,\,0.5,\,0.8$.

As mentioned in paper\,II, rotation has two major effects on the evolutionary
tracks.

\begin{figure}[!htp]
\centering
\includegraphics{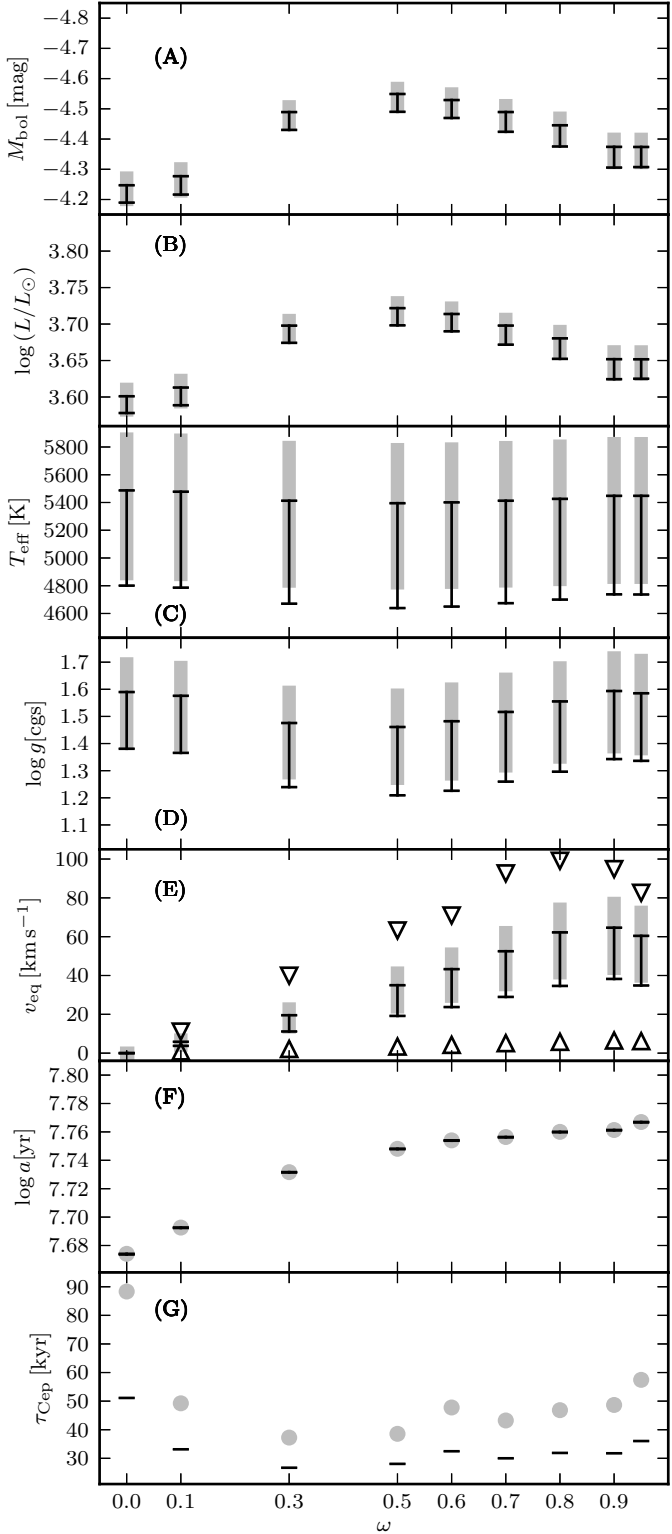}
\caption{Predictions for $7\,M_\odot$ Cepheids on the \emph{second crossing}
as function of $\omega$. Panels A through G show bolometric magnitude,
luminosity, effective Temperature, surface gravity, equatorial velocity,
age, and lifetime. The range of values during the IS crossing is 
shown as black lines for the IS definition by \citet{2000ApJ...529..293B}, and
as a light gray shade for the IS from \citet{2003A&A...404..423T}. On the second
crossing, all parameters increase during the evolution along the blue loop,
except for $\tau_{\rm{Cep}}$ which shows the 
duration of the IS crossing. Triangles in panel E show the extreme values for
surface velocity at the beginning of the blue loop near the Hayashi track
(minimal) and at its greatest extension (maximal).} 
\label{fig:observables_2C}
\end{figure}
\begin{figure}[!htp]
\centering
\includegraphics{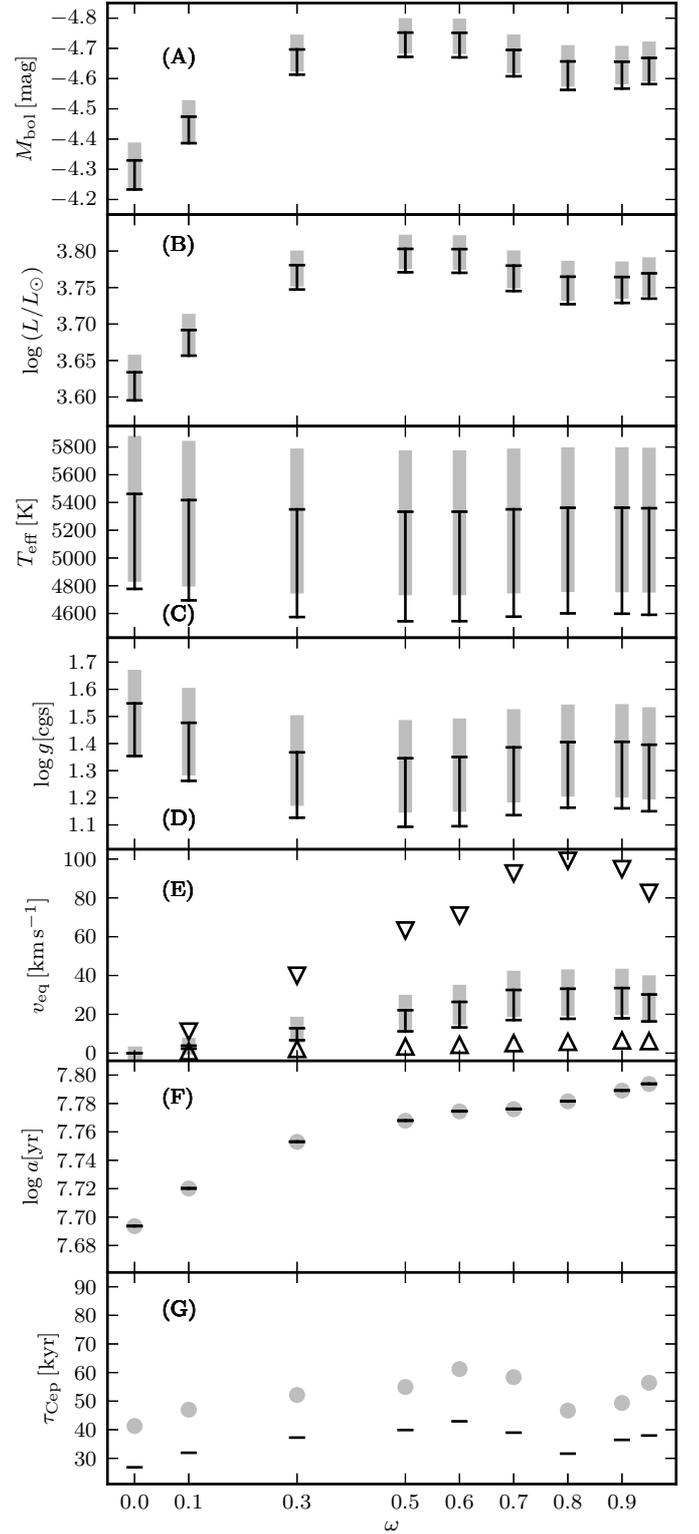}
\caption{Analogous to Fig.\,\ref{fig:observables_2C} for the \emph{third
crossing}. Note that the parameters in panels A through E are \emph{decreasing}
during the third  IS crossing.}.
\label{fig:observables_3C}
\end{figure}

Firstly, rotational mixing increases stellar core size and extends MS lifetimes,
mixing hydrogen from outer layers into the core.
It can be shown \citep[p.\,41, derived for the conditions close to the
center]{2009pfer.book.....M} that luminosity depends on mass, mean molecular weight, $\mu$, and opacity, $\kappa$. Due to the additional mixing, a star in rotation can continue burning hydrogen for
longer and convert more hydrogen into helium, resulting in higher $\mu$, which
increases $L \propto \mu^4$. This furthermore reduces the (dominant) electron
scattering opacity, since $\kappa = 0.2 \cdot (1 + X)$ is lowered when the hydrogen mass
fraction $X$ is reduced, and results in increased luminosity, since $L \propto
1/\kappa$. 

Secondly, rotation causes the
centrifugal force to modify hydrostatic equilibrium, leading to lower effective
core mass, i.e., the star evolves as if it had lower mass, resulting in a
decrease of luminosity. Hence, the ZAMS luminosity
of a rotating model corresponds to a ZAMS luminosity of a non-rotating model of
lower effective mass. Conversely, a non-rotating model
requires higher mass to reach the same luminosity as a rotating  model. Figure
\ref{fig:tracks} serves to illustrate this: whereas little difference in ZAMS
luminosity is seen between the non-rotating (blue solid line) and $\omega=0.5$
models (green dashed line), the ZAMS luminosity of the $\omega=0.8$ models (red
dash-dotted) is significantly lower. 

While the above is correct for a model at the ZAMS, the situation is very
different at later evolutionary stages, e.g. when a star crosses the IS.
Here, a rotating model has a luminosity that corresponds to a non-rotating model
of higher initial mass. The increase in luminosity due to
rotation is, however, not montonous, since  
the luminosity-increasing mixing effects are counteracted by 
the luminosity-decreasing centrifugal force.

The B panels in Figs.
\ref{fig:observables_2C} and\,\ref{fig:observables_3C} show this balancing effect 
for the $7\,M_\odot$ models for the full range of initial rotation rates
available from paper\,II.
Luminosity reaches a maximum for 
$\omega \approx 0.5$ and decreases towards both extremes, with a relatively flat
plateau for $0.3 < \omega < 0.7$. On the high $\omega$ end,
the decrease in luminosity is greater during the second crossing than during
the third, i.e. the blue loop becomes wider with greater rotation. Hence, the
faster $\omega$, the greater the difference in $L$ between second and third
crossing.

\subsection{Cepheid masses}\label{sec:CepMass}
Let us now consider the effect of rotation on the mass predictions for
Cepheids. We first inspect the upper and lower mass limits of (second and third
crossing) Cepheids predicted by the models and then investigate the impact of
rotation on the mass discrepancy problem.

\subsubsection{Which progenitor masses become Cepheids?}
As mentioned above, we assume that any model performing a blue loop
becomes a Cepheid when it crosses the IS. 
Within this simplification, our models allow an investigation of the range of
initial masses that evolve to become Cepheids, i.e., that develop blue loops
crossing the IS, as a function of rotation. At the low mass end, this inspection
can be done via interpolation, since the extent of the blue loops increases
gradually with mass. At the upper mass end, however, the blue loop suddenly
disappears, rendering interpolation inapplicable to this end. Depending on the
desired resolution in mass, this endeavor becomes computationally expensive
quickly. We therefore computed several new models to iteratively search for
the point at which the blue loop disappears\footnote{Rotating models calculated
here as in paper\,I, where rotation is parametrized as $v / v_{\rm{crit}}$, with $v_{\rm{crit}}$ denoting critical
rotation. $v / v_{\rm{crit}} = 0.4$ is equivalent to $\omega = 0.568$.}. 

Table\,\ref{tab:CepMass} presents the results of this search. Lower mass limits
are shown on the left, upper mass limits on the right. The next model
investigated is indicated in parenthesis. 

\begin{table}[t]
\centering
\begin{tabular}{@{}lr@{}}
\begin{tabular}{@{}l|r@{}}
lower mass limit & $v / v_{\rm{crit}}$  \\
\hline
\rule[0mm]{0mm}{3.5mm}$4.50\,M_\odot$\hspace{0.3cm}
(4.25) & 0.0 \\
\rule[0mm]{0mm}{3.5mm}$4.55\,M_\odot$\hspace{0.3cm} (4.50) &
0.4 \\ 
\hline
\end{tabular}
\hspace{0.5cm}
\begin{tabular}{@{}l|r@{}}
upper mass limit & $v / v_{\rm{crit}}$  \\
\hline
\rule[0mm]{0mm}{3.5mm}$11.50\,M_\odot$\hspace{0.3cm} (11.75) & 0.0 \\
\rule[0mm]{0mm}{3.5mm}$10.00\,M_\odot$\hspace{0.3cm} (10.25) &
0.4\\ 
\hline
\end{tabular}
\end{tabular}
\caption{Mass limits for Cepheids without and with rotation. Lower limits
(left) were explored using the web-based track interpolation tool. To
investigate the upper limits, new evolutionary tracks were computed. The values
in parentheses correspond to the next mass explored that no longer exhibits
a blue loop.}
\label{tab:CepMass}
\end{table}

On the low-mass end, we find that both rotating and non-rotating models 
yield a minimal Cepheid mass $\sim 4.5\,M_\odot$, independent of $\omega$.
Judging from Fig.\,\ref{fig:litModelComparison}, this is comparable to or lower
than the mass limits predicted by other models.
 
Rotation does, however, significantly affect the upper mass limit, and yields 
$M\,\lesssim\,11.5\,M_\odot$ for non-rotating, 
and $M\,\lesssim\,10.0\,M_\odot$ for rotating models with $v/v_{\rm{crit}} =
0.4$.
It is worth noting that the luminosity during the loop is
$\log{L/L_\odot}\approx 4.3$ for either case. 
Hence, the upper mass limit for Cepheids to exhibit blue loops appears to be
imposed by an upper limit on luminosity. Since rotation affects luminosity,
this affects the predicted Cepheid mass ranges. \emph{This leads to the 
interesting prediction that the longest-period Cepheids (the most
massive ones) in sufficiently large stellar populations may preferentially
have slowly rotating progenitors.}

\subsubsection{Evolutionary masses of rotating Cepheids}
A long-standing problem in Cepheid research is related to the Cepheid mass
anomalies \citep[see][and references therein]{1980ARA&A..18...15C}, which are
defined as disagreements between pulsational masses and those inferred from
evolutionary models
\citep{1968QJRAS...9...13C,1969MNRAS.144..461S,1969MNRAS.144..485S,1969MNRAS.144..511S}.
The improvement of radiative opacities in stellar envelopes
\citep{1991ApJ...371..408I,1994MNRAS.266..805S,1996ApJ...464..943I} removed some
of these anomalies, notably for the double-mode Cepheids
\citep{1992ApJ...385..685M}. For single-mode Cepheids, however, the so-called
mass discrepancy remains a topic of active research and discussion
\citep[e.g.][]{2006MmSAI..77..207B,2008ApJ...677..483K,2011ApJ...728L..43C,2012ApJ...749..108P}.
Evolutionary masses, i.e., those inferred from evolutionary models via a
mass-luminosity relationship, are found to be systematically larger than mass
estimates obtained by other means.

The most common strategies explored to resolve the mass discrepancy 
involve an increase in the size of the convective core via enhanced
overshooting \citep[e.g.][]{2008ApJ...677..483K}, a decrease in envelope mass
by enhanced mass loss \citep{2011A&A...529L...9N}, or
both. For instance, \citet{2012ApJ...749..108P} have recently
found that the accurately determined mass of the Cepheid component
in the eclipsing binary OGLE-LMC-CEP-0227 \citep{2010Natur.468..542P}
located in the Large Magellanic Cloud can be satisfactorily reproduced by
evolutionary models with a given set of parameters.
Their best-fit solution favors increased core size over enhanced mass-loss, but
requires a significantly higher amount of core overshooting than is implemented
in the models presented here.

Another possibility for increasing core size is to introduce rotation. However,
the evolutionary effect of rotation on Cepheids has not yet been discussed in
detail in the literature. This is unfortunate, since rotation makes testable
predictions for a range of parameters that can be confronted to observation,
including enhanced surface abundances and (equatorial) rotational velocities,
which make this effect distinguishable from enhanced core overshooting.
It is therefore clear that rotation should not be neglected as a potential contributor
to the solution of the mass discrepancy.

As mentioned in Sec.\,\ref{sec:HRD}, the luminosity of rotating
Cepheid models is generally larger than that of non-rotating Cepheids of the
same mass.
Judging from the tracks in Fig.\,\ref{fig:tracks}, for instance, the
increase in luminosity at fixed mass is approximately $50\%$ ($\approx 0.17$ in
$\log{L/L_\odot}$). A non-rotating Cepheid model must therefore
have higher mass than a rotating one at a fixed luminosity. 

\begin{figure}
\centering
\includegraphics{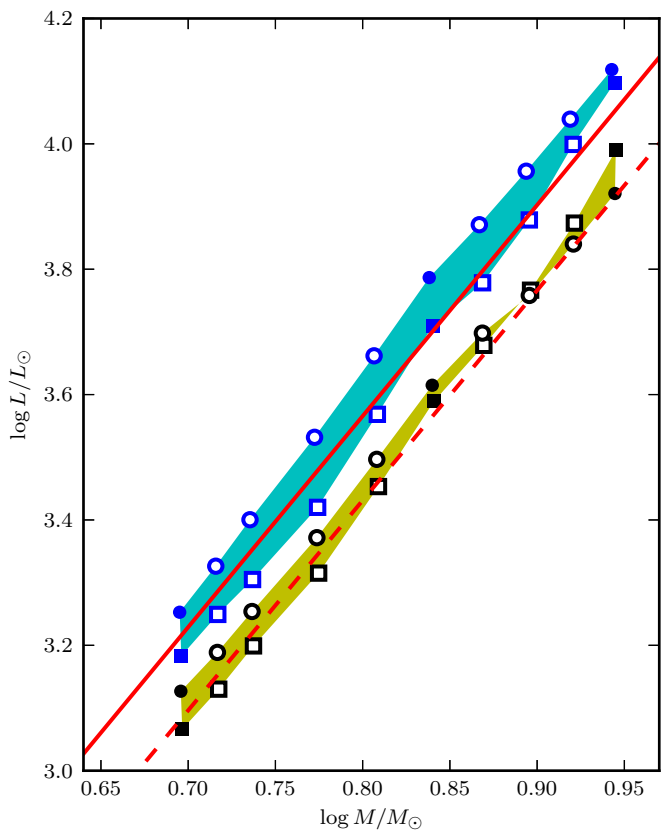}
\caption{The mass-luminosity relationship for rotating ($\omega=0.5$, cyan,
consistently higher luminosity) and non-rotating Cepheids (yellow, lower
luminosity).
Shaded areas show the range of values between the second (squares) and third
(circles) crossings. Solid symbols are based on grid models, open symbols
indicate models obtained via the web-based interpolation tool.
The red dashed line that crosses the non-rotating area represents the M-L
relation by \citet[their Eq. 2 with $Z=0.014$, $Y=0.27$]{2005ApJ...629.1021C}
whose investigation was based on the models by
\citet{2000ApJ...543..955B} and does not account for convective core
overshooting.
The solid red line crossing through the area delineated by rotating models
is the M-L relationship used by \citet{2013AJ....146...93R}, which assumes
a value of $d_{\rm{over}}/H_P = 0.2$ for convective core overshooting
\citep[based on models by][]{2012ApJ...749..108P}}.
\label{fig:MLR}
\end{figure}

Figure\,\ref{fig:MLR} serves to investigate the effect of rotation on the
mass-luminosity (M-L) relationship. It was created using the regular models from
paper\,II as well as interpolated models computed using the web-based
interpolation
tool\footnote{\url{http://obswww.unige.ch/Recherche/evoldb/index/Interpolation/}}.
As mentioned above, luminosity usually increases from the second to the third
crossing, and this increase tends to be larger for higher $\omega$, i.e., the
loops become wider for higher $\omega$.
Hence, it seems obvious to distinguish not only between models with different
$\omega$, but also between Cepheids on different crossings, since 
M-L relations of Cepheids on different crossings and
with different initial rotation rates have different zero-points and slopes.

Figure\,\ref{fig:MLR} shows both rotating (cyan shaded area) and
non-rotating M-L relations (yellow shaded area) based on the present models, as
well as literature M-L relations with different assumptions regarding convective
core overshooting: the \citet{2005ApJ...629.1021C} models
\citep[based on][]{2000ApJ...543..955B} assume no overshooting, whereas the
relation given in \citet{2013AJ....146...93R} assumes $d_{\rm{over}}/H_P = 0.2$
\citep[based on models by][]{2012ApJ...749..108P}. Thus, Fig.\,\ref{fig:MLR}
provides a means to compare the effects of increasing convective core
overshooting via the literature relations, as well as of introducing
rotation via the shaded areas (note, however, that the present models assume
weak overshooting with $d_{\rm{over}}/H_P = 0.1$).

Figure\,\ref{fig:MLR} clearly exposes two important aspects to bear in mind
with respect to the M-L relation.
First, rotation increases Cepheid luminosities at a similar rate as high
convective core overshooting values, thus pointing to the presence of a
degeneracy between the two effects in terms of the M-L relation.
Second, it is crucial to take into account the crossing number when inferring
Cepheid masses, and the importance of doing so increases with $\omega$ (since
the luminosity difference widens with $\omega$).
Hence, if rotation and crossing number are ignored, systematic errors will be
made when inferring masses from a given M-L relation. As can be seen from Fig.
\ref{fig:MLR}, evolutionary masses will be overestimated,
leading to a mass discrepancy.  

Let us compare the systematic error on mass estimates incurred 
in four different situations:
\begin{enumerate}
  \item both $\omega$ and the crossing number are unknown: using
  \emph{non-rotating} models, masses are overestimated by  $12-15\,\%$.
  \item $\omega$ is known, but not the crossing: assuming the star is on the
  second crossing while it is on the third, the masses are overestimated by
  2-7\,\%.
  \item The Cepheid is on the second crossing, $\omega$ is unknown: using
  \emph{non-rotating} models, masses are overestimated by  $8\,\%$.
  \item The Cepheid is on the third crossing, $\omega$ is unknown: using
  \emph{non-rotating} models, masses are overestimated by  $10-17,\%$.
\end{enumerate}

Note that these systematic offsets approach the range of
$10-20\,\%$ usually quoted for the Cepheid mass discrepancy
\citep[e.g.][]{2006MmSAI..77..207B,2008ApJ...677..483K}. The crucial
point to remember is that two effects need to be taken into
account simultaneously when inferring evolutionary masses: the rotational
history of the star ($\omega$), and the crossing number.
Thankfully, both can in principle be constrained by observations, the former
via estimates of $v\sin{i}$, Cepheid radii, and surface abundance enrichment,
and the latter via rates of period changes \citep[e.g.][]{2006PASP..118..410T}. 

Due to the balance between the mixing and hydrostatic effects, the majority
of Cepheid models predict luminosities that deviate not too far from
the $\omega=0.5$ models, cf. Figs.\,\ref{fig:observables_2C} and
\ref{fig:observables_3C}. 
M-L relations for Cepheids based on rotating models with $\omega=0.5$ therefore
provide suitable estimates for a range of initial rotation rates observed in
B-type stars on the MS. 

The rotation-averaged M-L relation for the \emph{second crossing} is thus:
\begin{equation}
\log{(L/L_\odot)} = (3.683 \pm 0.074)\log{(M/M_\odot) + (0.598 \pm 0.006)}
\label{eq:LMRsecond}
\end{equation}
\begin{equation}
\log{(M/M_\odot)} = (0.271 \pm 0.006)\log{(L/L_\odot)} - (0.159 \pm 0.001)\,.
\label{eq:MLRsecond}
\end{equation}
Analogously for the \emph{third crossing}, we obtain:
\begin{equation}
\log{(L/L_\odot)} = (3.515 \pm 0.043)\log{(M/M_\odot) + (0.818 \pm 0.004) }
\label{eq:LMRthird}
\end{equation}
\begin{equation}
\log{(M/M_\odot)} = (0.284 \pm 0.003)\log{(L/L_\odot)} - (0.231 \pm 0.001)\,.
\label{eq:MLRthird}
\end{equation}
If the crossing number is also unknown, then an average relation in between
these two should be used. We thus propose
\begin{equation}
\log{(L/L_\odot)} = (3.594 \pm 0.124)\log{(M/M_\odot) + (0.712 \pm 0.011)}
\label{eq:LMRnone}
\end{equation}
\begin{equation}
\log{(M/M_\odot)} = (0.272 \pm 0.009)\log{(L/L_\odot)} - (0.177 \pm 0.002)
\label{eq:MLRnone}
\end{equation}
to be used as an M-L relation in the absence of information on $\omega$ and the
crossing number. We note that this average relation yields masses for
low-luminosity ($\log{L/L_\odot} \approx 3.2$) Cepheids that agree to within
less than $1\%$ with the relationship by \citet[non-canonical overshooting]{2012ApJ...749..108P} as given in \citet{2013AJ....146...93R}. Towards higher luminosity, the difference between
the relationships increases, reaching nearly $5\%$ at $9\,M_\odot$ (rotating
models predict lower mass).

\begin{figure}
\includegraphics{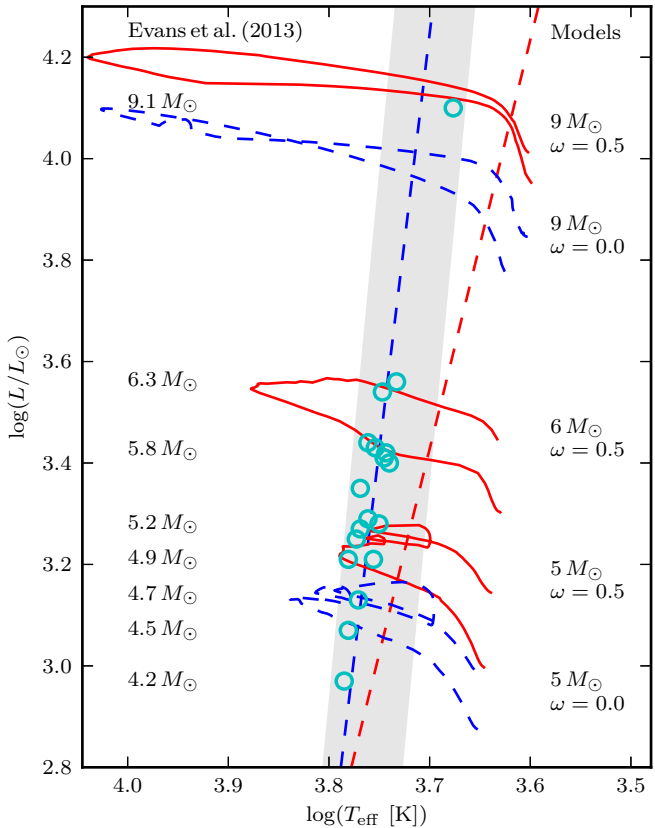}
\caption{Cepheid with masses determined by \citet{2013AJ....146...93R}
plotted as cyan open circles onto the blue loop portions of evolutionary tracks.
Rotating tracks are drawn as red solid, non-rotating ones as blue dashed lines,
the IS as in Fig.\,\ref{fig:tracks}.}
\label{fig:EvansHRD}
\end{figure}

Figure\,\ref{fig:EvansHRD} illustrates the importance of considering the crossing
number when inferring the mass of a Cepheid. Cepheids with masses listed in
\citet{2013AJ....146...93R} are plotted as cyan open circles over the blue loop
portions of evolutionary tracks in an HRD. Cepheid mass, luminosity and color
are provided by \citet{2013AJ....146...93R}, and the temperature estimate is
obtained by interpolating in the color grid by \citet{2011ApJS..193....1W}  
assuming solar metallicity and $\log{g}=1.5$.

The figure clearly shows that there is overall very good agreement between the
literature masses and the models. Furthermore, it underlines the penalty of
ignoring the crossing number when inferring Cepheid masses. For instance, if
the crossing is ignored, then Cepheids between $5.8$ and $6.3\,M_\odot$ may both
be interpreted as $6\,M_\odot$ Cepheids. Figure\,\ref{fig:EvansHRD}
also illustrates that the present models predict luminosities that are
consistent with the period-luminosity relation (PLR) by
\citet{2007AJ....133.1810B}, since the Cepheid luminosities in
\citet{2013AJ....146...93R} are based on this PLR relation together with an M-L
relation which is analytically similar to Eq.\,\ref{eq:MLRnone}.

\begin{figure}
\centering
\includegraphics{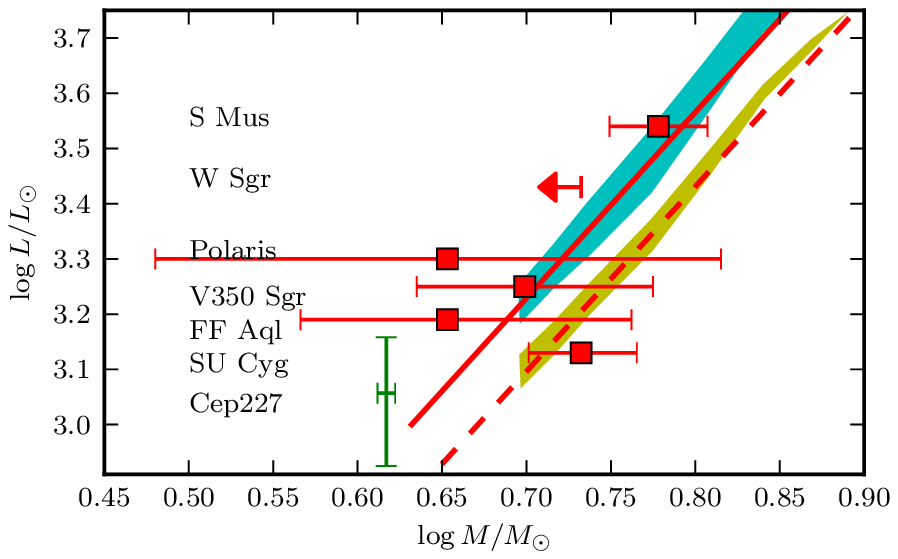}
\caption{Observed (non-model dependent) Cepheid masses
\citep[see][and references therein]{2011AJ....142...87E} shown together with the
M-L relations of Fig.\,\ref{fig:MLR}.
Luminosities were estimated by \citet{2013AJ....146...93R} or 
\citet[Polaris \& FF Aql]{2011AJ....142...87E}
using the period-luminosity relationship by
\citet{2007AJ....133.1810B}. The green errorbars labeled `Cep227' show the
highly accurate mass estimate for OGLE-CEP-LMC-0227 \citep{2010Natur.468..542P},
where the luminosity was calculated from the published radius and temperature.}
\label{fig:MLRobs}
\end{figure}
Few model-independent Cepheid masses are known in the literature, and
unfortunately, their associated uncertainties are rather large. In Fig.
\ref{fig:MLRobs}, we compare the model-independent masses shown in
\citet{2011AJ....142...87E} with the M-L relations presented in Fig.
\ref{fig:MLR} above. This comparison is not very conclusive, unfortunately,
though it does seem to favor the rotating M-L region shown. An interesting test
to perform would be to investigate the Cepheid SU\,Cyg in terms of its
observables related to rotation, since its location in the diagram is only
consistent with virtually no rotation. Note that M-L relations are
sensitive to metallicity in the sense that lower metallicity models yield
higher luminosity. This is important to keep in mind when considering  
the LMC Cepheid OGLE-LMC-CEP-0227 \citep{2010Natur.468..542P} shown by green
error bars.

While there does exist a degeneracy between the adopted value for the
overshooting parameter and the rotation rate in terms of the M-L relation,
rotation has implications on a star's evolution that may be observable in the
late stages of its evolution. For instance, rotation leads to enhanced helium
surface abundance and modified CNO element abundance ratios, see
Sec.\,\ref{sec:PropertiesSurfAbund}. 
For Cepheids of a fixed mass, enhanced helium abundance would increase
pulsation amplitudes, since the total amount of helium in the partial ionization zone
would be increased, see Sec.\,\ref{sec:enricHe}. 
Furthermore, rotating models predict surface velocities that are clearly within
the detectable range, cf. Sec.\,\ref{sec:veq}.
Hence, the present rotating models make several potentially observable
predictions that can be used to constrain the value of $\omega$ for a given
Cepheid, and help to distinguish between evolutionary effects due to rotation
and those due to higher core overshooting.

To summarize the above, the current mass discrepancy (in the
order of $10-20\%$) can be explained by a combination of increased luminosity
due to rotation and luminosity differences between Cepheids on second and
third IS crossings.

\subsection{Surface gravity, radii, and equatorial velocity}\label{sec:veq}
As can be seen in the D panels of Figs.\,\ref{fig:observables_2C}
and\,\ref{fig:observables_3C}, rotation impacts surface gravity, $\log{g}$,
during the IS crossings. Qualitatively similar in behavior to the increase in
luminosity due to rotation, rotating models tend to have lower $\log{g}$, and
the range of $\log{g}$ values between the two crossings increases with rotation.
This is no surprise, since $\log{g}$ and $L$ are intimately linked via the
M-L relationship. Since $\log{g}$ is related to stellar radius, rotation also
impacts a Cepheid's radius. These effects are particularly interesting to
mention here, since $\log{g}$ and radius can be determined observationally and
independently, providing constraints on a Cepheid's crossing number and on
$\omega$.

Due to the conservation of angular momentum, stars with greater rotation rates
on the MS can be expected to have greater rotational velocities also during the
red (super-) giant phase. Panels E of Figs.\,\ref{fig:observables_2C} and  
\ref{fig:observables_3C} show the dependence of the equatorial surface velocity
for the $7\,M_\odot$ models, $v_{\rm{eq}}$ on initial rotation rate during the
second and third crossings: $v_{\rm{eq}}$ increases during the second  
crossing, decreases during the third. The minimum and maximum values during the
entire blue loop are indicated by up- and downward facing triangles,
respectively.

\begin{table}
\centering
\begin{tabular}{@{}ll|rr|rr|rr|@{}}
\multicolumn{2}{l}{Models} & 
\multicolumn{2}{|c}{first}&\multicolumn{2}{c}{second}&\multicolumn{2}{c|}{third}
\\
$M/M_\odot$ & $\omega$ & BE & RE & RE & BE & BE & RE \\ \hline 
\rule[0mm]{0mm}{3.5mm}5.0 & 0.5 & 23.9 & 21.8 & 20.8 & 24.2 & 14.2 &  7.6 \\
\rule[0mm]{0mm}{3.5mm}    &     & 26.1 & 17.6 & 18.5 & 25.4 & 21.2 & 5.8 \\  
\hline
\rule[0mm]{0mm}{3.5mm}7.0 & 0.5 & 14.6 & 11.0 & 19.2 & 35.0 & 22.1 & 11.2 \\
\rule[0mm]{0mm}{3.5mm}    &     & 16.6 & 10.8 & 23.8 & 41.2 & 26.6 & 15.2 \\
 \hline
\rule[0mm]{0mm}{3.5mm}9.0 & 0.5 & 9.8  &  6.6 &  8.1 & 37.6 & 17.7 & 4.6 \\ 
\rule[0mm]{0mm}{3.5mm}    &     & 11.6 &  7.7 & 22.7 & 52.0 & 23.8 & 11.5 \\
\hline
\end{tabular}
\caption{Range of equatorial velocities $v_{\rm{eq}}$ in
$\rm{km\,s^{-1}}$ predicted by models with $\omega=0.5$ for three different
masses upon entering and exiting the blue (BE) and red (RE) edges of the IS.
For each mass, values based on the \citet{2000ApJ...529..293B} IS are presented
first, followed by those for the \citet{2003A&A...404..423T} IS.}
\label{tab:veq_crossings}
\end{table}
Equatorial velocities predicted for models with $\omega=0.5$
for $5,\,7,\,$ and $9\,M_\odot$ models between the red (RE) and blue (BE) edges 
of the IS are listed in Tab.\,\ref{tab:veq_crossings}.
$v_{\rm{eq}}$ tends to be consistently larger on the second crossing than on the
third due to larger radius (lower $\log{g}$). 
Given the predictions in Tab.\,\ref{tab:veq_crossings}, the presence of high
$v\sin{i}$ would tend to be indicative of a Cepheid being on the second
crossing. 

At the velocities predicted here, rotation should be readily observed by
spectral line broadening.
A first estimate of the average $v_{\rm{eq}}$ based on a sample of $97$
classical Cepheids observed with the high-resolution spectrograph {\it Coralie} as part of
an observing program to search for Cepheids associated with open clusters
\citep{2013MNRAS.434.2238A} yields $\langle v_{\rm{eq}} \rangle \approx
12.3\,\rm{km\,s^{-1}}$ (R.~I. Anderson 2013, PhD thesis). A detailed comparison
with observed surface velocities is currently in preparation. 
Three published observational studies found $v_{\rm{eq}} \lesssim
10\,\rm{km\,s^{-1}}$ \citep[using CORAVEL]{1996A&A...306..417B}, $v\sin{i}
\leq 16\,\rm{km\,s^{-1}}$ \citep[estimates of $v \sin{i}$ in their
Tab.\,2]{2006A&A...453..309N}, and $4.9 \leq v\sin{i} \leq
17.7\,\rm{km\,s^{-1}}$ \citep[Cepheids identified via
cross-match with the Variable Star Index, cf. 
\citealt{2006SASS...25...47W}]{2013arXiv1312.3474D}.

At first sight, this comparison may appear to
indicate overestimated predictions for $v_{\rm{eq}}$ for models with initial
$\omega=0.5$.
Note, however, that the predicted values of $v_{\rm{eq}}$ vary significantly
between the different crossings, within each crossing, as a function of mass (or
period), and are moreover strongly dependent on the choice of instability strip
(bluer boundary = more contracted = higher $v_{\rm{eq}}$). A detailed comparison
must therefore take into account the crossings, temperature, masses (periods),
and be based on a sufficiently large sample in order to deal with the randomly oriented
rotation axes.

\subsection{Cepheid ages \& lifetimes}\label{sec:lifetimes}
\begin{table}[t]
\centering
\begin{tabular}{@{}l|c|c|c|c|c|c|@{}}
 & \multicolumn{6}{|c|}{$\log{a}$ [yr]} \\
 & \multicolumn{3}{|c|}{second crossing}  & \multicolumn{3}{c|}{third crossing}
 \\ \hline 
\rule[0mm]{0mm}{3.5mm}$\omega$ & 0.0 & 0.5 & 0.8 & 0.0 & 0.5 & 0.8 \\
\hline
\rule[0mm]{0mm}{3.5mm}$5\,M_\odot$ &  8.01 & 8.09  & 8.11& 8.03   & 8.10  & 8.12 
\\
\rule[0mm]{0mm}{3.5mm}$7\,M_\odot$ & 7.67  & 7.75  & 7.76 & 7.69 & 7.77 & 7.78 
\\
\rule[0mm]{0mm}{3.5mm}$9\,M_\odot$ & 7.46  & 7.53  & 7.54* & 7.48 & 7.55 & 7.56* 
\\ \hline
\end{tabular}
\caption{Cepheid ages for three different masses and rotation
rates. The typical difference between rotating and non-rotating models is
approximately $20\%$ across the board. *:
for $9\,M_\odot$ models, this column actually shows $\omega=0.7$.}
\label{tab:Cepages}
\end{table}
Rotating models yield older Cepheids than non-rotating ones (see panels F in
Figs.\,\ref{fig:observables_2C} and \ref{fig:observables_3C}), since rotation
increases the MS lifetime of stars.
For example, the difference in mean age during the second crossing between  
the non-rotating $7\,M_\odot$ model ($49.4$\,Myr), and the $\omega=0.5$
model of the same mass ($58.6$\,Myr) is approximately $20\%$.
Additional ages for Cepheid models of $5$, $7$, and $9\,M_\odot$ models for
different initial rotation rates are listed in  Tab.\,\ref{tab:Cepages}.
It is evident from the table that moderate rotation ($\omega=0.5$) causes a
systematic increase in $\log{a}$ 
of approximately $0.08$\,dex, regardless of mass. Hence,
the present models suggest that (rotating) real Cepheids may be systematically
older by $20\%$ than predicted by non-rotating models.
Such considerations are relevant for calibrations of
period-age relations \citep{2005ApJ...621..966B} and their applications for    
constraining star formation histories, e.g. of the Galactic
nuclear bulge \citep{2011Natur.477..188M}.

\begin{table*}
\centering
\begin{tabular}{@{}l|r@{\hskip 3.5mm}r@{\hskip 3.5mm}r|r@{\hskip
3.5mm}r@{\hskip 3.5mm}r|r@{\hskip 3.5mm}r@{\hskip 3.5mm}r|r@{\hskip
3.5mm}r@{\hskip 3.5mm}r|r@{\hskip 3.5mm}r@{\hskip 3.5mm}r|r@{\hskip
3.5mm}r@{\hskip 3.5mm}r|@{}} & \multicolumn{12}{|c|}{$\tau_{\rm{Cep}}$ [kyr]} &
\\
 & \multicolumn{3}{|c|}{first} & \multicolumn{3}{|c|}{second} &
 \multicolumn{3}{|c|}{third} & \multicolumn{3}{|c|}{total} &
 \multicolumn{3}{|c|}{$\tau_{\rm{loop}}$ [Myr]}  &
  \multicolumn{3}{|c|}{$\tau_{\rm{He}}$ [Myr]}  \\ \hline
\rule[0mm]{0mm}{3.5mm}$\omega$ & 0.0 & 0.5 & 0.8 & 0.0 & 0.5 & 0.8 & 0.0 & 0.5 &
0.8 & 0.0 & 0.5 & 0.8 & 0.0 & 0.5 & 0.8 & 0.0 & 0.5 & 0.8 \\
\hline
\rule[0mm]{0mm}{3.5mm}$5\,M_\odot$  & 5.7 & 9.4 & 8.5 & 316 & 288 & 857 & 597 &
1553 & 1145 & 919 & 1851 & 2011 & 10.3 & 8.4 & 9.3 & 19.2 & 16.6 & 18.7 \\
\rule[0mm]{0mm}{3.5mm} & 38 & 35 & 35 & 879 & 930 & 2681 & 1977 & 3422 & 2804 &
 2894 & 4387 & 5520 & & & & & & \\ \hline
\rule[0mm]{0mm}{3.5mm}$7\,M_\odot$  & 4.0 & 4.5 & 4.1 & 51 & 28 & 32 & 27 & 40 &
31.6 & 82 & 72 & 68  & 3.0  & 3.4 & 3.9 & 6.8  & 6.7  & 7.3 \\
\rule[0mm]{0mm}{3.5mm} & 8.0 & 7.4 & 7.2 & 88 & 39 & 47 & 41 & 55 & 47 & 138 &
 101 & 101 & & & & & & \\ \hline 
\rule[0mm]{0mm}{3.5mm}$9\,M_\odot$* & 2.2 & 3.0 & 2.7 & 14 & 11 & 11 & 6 & 13 &
10 & 22 & 27 & 24    & 1.5 & 1.6 & 1.9 & 3.4 & 3.3 & 3.6 \\
\rule[0mm]{0mm}{3.5mm} & 2.7 & 2.7 & 2.6 & 9.4 & 6.7 & 8.9 & 6.9 & 12 & 9.2 & 19
 & 21 & 21         & & & & & & \\
\hline 
\end{tabular}
\caption{Predicted timescales relevant for Cepheids of different initial mass
and rotation rate. Cepheid lifetimes in the three crossings and their sum
(total lifetime) are listed in [kyr]. The duration of the blue loop and the
core helium burning timescale are given in [Myr].
The asterisk is a reminder that $\omega=0.7$ was used instead of $\omega=0.8$
for $9\,M_\odot$ models.
The predicted values in the top rows for each mass are based on
the \citet{2000ApJ...529..293B} IS, the lower rows on
\citet{2003A&A...404..423T}.} 
\label{tab:Ceplifetimes}
\end{table*}

Besides age, rotation also impacts the duration of blue loops, and therefore 
Cepheid lifetimes (the time spent crossing the IS). Cepheid lifetimes and
blue loop durations are listed in kyr in Tab.\,\ref{tab:Ceplifetimes} covering three
rotation rates with three initial masses for both the
\citet{2000ApJ...529..293B} and \citet{2003A&A...404..423T} instability strips.
We note here that the $5\,M_\odot$ models are affected
by helium spikes during the third crossing.
These occasional sudden increases of available helium can extend the lifetime of
the third crossing significantly, if they occur inside the IS. However, 
\emph{Cepheid lifetime estimates depend much more strongly on the
definition used for the IS boundaries} than on these He spikes.
Hence, the reader is advised to consider the following discussion merely as an
indicator of tendencies. 

The overall trends predicted for Cepheid lifetimes in Tab.
\ref{tab:Ceplifetimes} are as follows:
\begin{itemize}
  \item Lifetime estimates for the $5$ and $7\,M_\odot$ models are much
  longer (by a factor 2-3 in the case of $5\,M_\odot$) when adopting the
  \citet{2003A&A...404..423T} IS rather than the \citet{2000ApJ...529..293B} IS.
  The longer lifetimes can be explained by the greater width
  (in $T_{\rm{eff}}$) of the former IS definition and its bluer hot edge
  (evolution slower on the blue edge than the red). 
  For $9\,M_\odot$ models, the \citet{2000ApJ...529..293B} IS predicts longer
  lifetimes due to its wedge shape.  
  \item First crossings are much faster than second or third crossings.
  The first crossing lifetime increases towards lower masses.
  \item The larger the mass, the longer the time spent on the first
  crossing, relative to the second or third. Hence, the larger the mass, the
  more probable it is for a Cepheid to be observed during the first crossing.
  \item Cepheid lifetimes for the $5\,M_\odot$ model are by far the longest.
  Compared to $7\,M_\odot$ models, $5\,M_\odot$ models predict lifetimes that
  are at least an order of magnitude longer, depending strongly on the 
  IS boundaries adopted and on rotation. The reason is that evolution along the
  blue loop is slowest at the tip of the blue loop, which lies square inside the IS for these models.
  \item For the $5\,M_\odot$ model, the third crossing is always slower than
  the second. However, this is not always the case for the higher-mass models.
  \item The fractional time spent inside the IS,
  $\tau_{\rm{Cep,total}}/\tau_{\rm{loop}}$, decreases with increasing mass. This
  is related to the speed at which a star evolves along the IS. As a result, 
  a $5\,M_\odot$ red giant is more likely to be caught (observed) during the
  Cepheid stage than an ordinary red giant of $7$ or $9\,M_\odot$. This is in
  addition to the effect of the IMF.
  \item The total lifetimes of intermediate and high mass Cepheids depend
  less on rotation than low-mass Cepheids do.
\end{itemize}

Better resolution in rotation rate regarding Cepheid lifetimes is provided
for the $7\,M_\odot$ model in the G panels of
Figs.\,\ref{fig:observables_2C} and \ref{fig:observables_3C}, though no clear trend is discernible.
  
\subsection{Surface abundance enrichment}\label{sec:PropertiesSurfAbund}

Rotational mixing creates effects that can impact Cepheid surface abundances
during two different evolutionary stages. First, during the MS, rotational
shear creates turbulence that slowly mixes material processed in the core 
throughout the radiative envelope, leading to
enhanced surface abundances of helium and of nitrogen relative to
carbon and oxygen. 
When the star evolves to become a red giant or supergiant, the core material is
carried to the stellar surface during the first dredge-up phase when the star develops a deep convective envelope during
its approach of the Hayashi track after the first IS crossing. Hence, two kinds
of surface abundance enhancement have to be distinguished: the
rotational enhancement occurring towards the end of the Main
Sequence, and the dredge-up related enhancement.
As the predictions presented in the following clearly show, both kinds of
surface abundance enhancement depend on mass and rotation.

\begin{figure}
\centering
\includegraphics{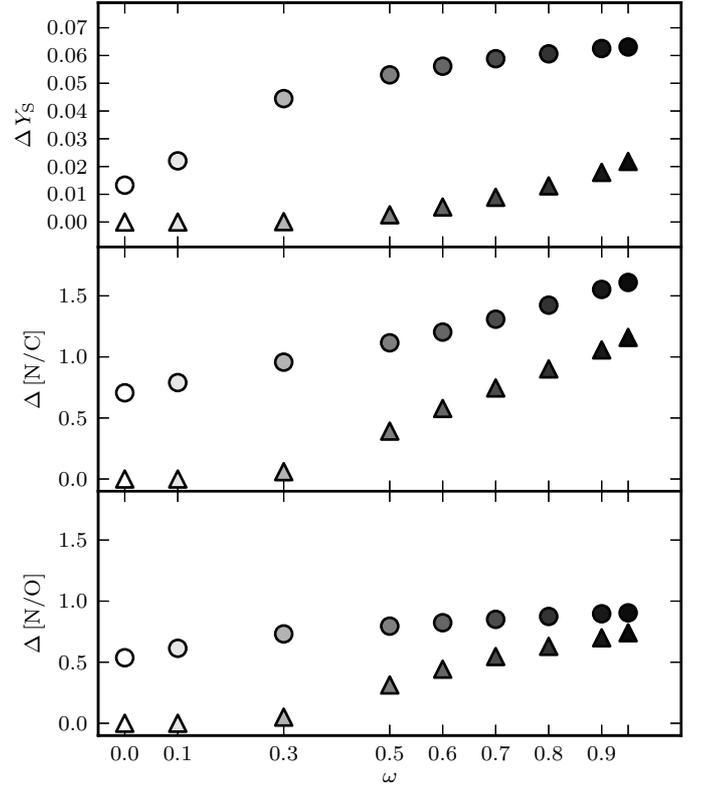}
\caption{Surface abundance enrichment for the $7\,M_\odot$
model as a function of initial rotation. Top panel shows surface helium mass
fraction relative to starting value, center and bottom panels show [N/C] and
[N/O] enhancement relative to starting (solar) value. Quantities before first
dredge up is shown as upward triangles, post-dredge up as circles.}
\label{fig:RotEnhancement}
\end{figure}

Abundance enhancement of element $X_{\rm{i}}$ is defined as the
post-dredge-up (index Cep) increase in mass fraction relative to the
initial value (on the ZAMS, index 0), i.e.,
\begin{equation}
\Delta X_{\rm{i}} = \Delta X_{\rm{i,DU}} + \Delta X_{\rm{i,MS}} \equiv
X_{\rm{i,Cep}} - X_{\rm{i,0}} \equiv X_{\rm{i,Cep}} - X_{\rm{i,\odot}} \, .
\end{equation}
$\Delta X_{\rm{i}}$ has two contributions, one due to the dredge-up (index DU),
and one related to rotational enhancement on the MS (index MS). 

For CNO elements, we express the abundance enhancement
in logarithmic units relative to hydrogen, where, in the case of nitrogen (N):
\begin{equation}
\rm{[N/H]} = \log{ \left( \frac{ X_{\rm{N,Cep}} }{
\rm{A}_{\rm{N}} \cdot X_{\rm{H,Cep}}  } \right) } - 
\log{ \left( \frac{ X_{\rm{N,\odot}} }{\rm{A}_{\rm{N}} \cdot X_{\rm{H,\odot}}  }
\right) } \equiv \Delta \rm{[N/H]} \, .
\end{equation} 
Here, $\rm{A}_{\rm{N}}$ denotes the atomic mass number of nitrogen.
Similarly, for ratio of nitrogen to carbon,
\begin{equation}
\Delta \rm{[N/C]} \equiv \rm{[N/C]} \equiv \rm{[N/H]} - \rm{[C/H]} \, .
\end{equation}

\begin{figure}[!h]
\centering
\includegraphics{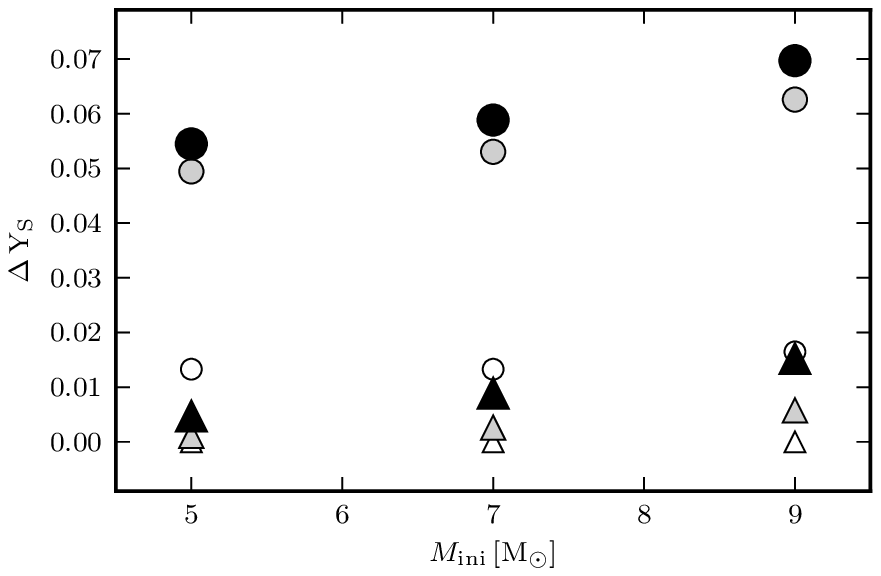}
\caption{Helium surface mass fraction enhancement as a function of mass
and initial rotation. Open symbols represent non-rotating models, gray solid symbols
$\omega=0.5$, and black solid symbols $\omega=0.8$.
Pre-dredge-up enhancement is drawn as triangles, post-dredge-up values as
circles.}
\vspace{0.5cm}
\label{fig:HEenhMass}
\includegraphics{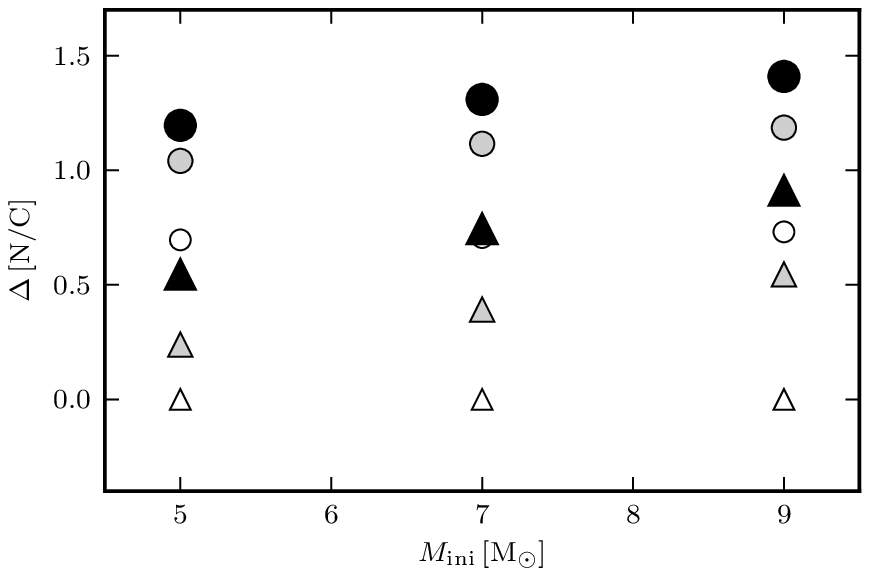}
\caption{Analogous to Fig.\,\ref{fig:HEenhMass} for [N/C] surface abundance
enhancement.}
\label{fig:NCenhMass}
\vspace{0.5cm}
\includegraphics{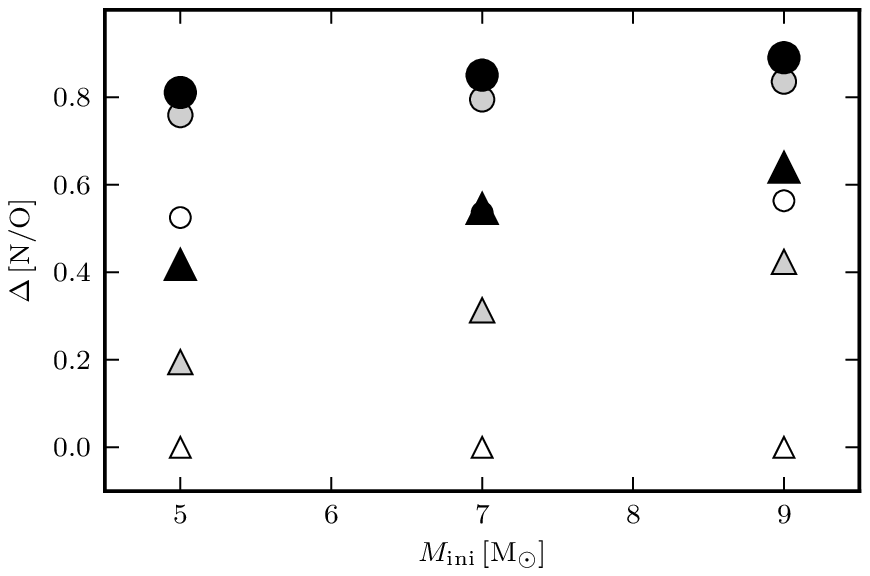}
\caption{Analogous to Fig.\,\ref{fig:HEenhMass} for [N/O] surface
abundance enhancement.}
\label{fig:NOenhMass}
\end{figure}

Figure\,\ref{fig:RotEnhancement} shows predicted surface abundance enhancements
for helium, $\Delta \rm{Y_S}$, as well as N relative to C and N relative to O as
a function of initial rotation rate for $7\,M_\odot$ models. 
Similarly, Figs.\,\ref{fig:HEenhMass},\,\ref{fig:NCenhMass}, and
\ref{fig:NOenhMass} show the same information as  
functions of both rotation (gray scaled, where darker represents higher
$\omega$) and mass.
In all these figures, Terminal Age Main Sequence (TAMS)
values are indicated as upward facing triangles. Post dredge-up values are shows
as solid circles. As per their definition, initial (ZAMS) values are 0.

Since rotational mixing can enhance surface abundances towards 
the TAMS, the presence of enriched abundances is not an unambiguous argument for
excluding a Cepheid's crossing the IS for the first time. As shown in the
following subsections, the TAMS enhancement depends primarily on rotation, but also
on mass. Thus, especially high mass, i.e., long-period, Cepheids are expected to
exhibit enhanced abundances even if they are on the first crossing.
However, an absence of enhanced abundances may be indicative of small
$\omega$ (rotational history of the Cepheid) and a first IS crossing. 

\subsubsection{Helium}\label{sec:enricHe}
Figure\,\ref{fig:HEenhMass} clearly shows that even small amounts of rotation
considerably increase the post-dredge-up helium surface mass fraction
$Y_{\rm{S}}$ drawn as circles whose recycle represents rotation rate (the
darker, the faster). High rotation rates ($\omega > 0.5$) are required to
cause significant enrichment on the MS (TAMS predictions drawn as upward
triangles with same grayscale), although enhancement during the MS is a factor
of 5-6 smaller than the post-dredge-up enhancement.

For a fixed mass, an increase in $Y_{\rm{S}}$ due to rotation also signifies an
increase in the total mass of helium in the He partial ionization zone, $\Delta
M_{\rm{He}}$. This should affect the pulsations, since the amplitude of light
variation may depend on He abundance. For instance,
\citet{1998A&AT...17...15K} predicted an increase in amplitude $\left( \delta L
\propto \Delta M_{\rm{He}} / P \right)$, where $P$ denotes pulsation period.
According to \citet{1980SSRv...27..419C}, an increase of He mass fraction by
$0.1$ would increase the light amplitude by $0.25$\,mag, and radial velocity
amplitude by $15\,\rm{m\,s^{-1}}$. However, in a recent investigation using
nonlinear convective hydrodynamical models, \citet{2013ApJ...768L...6M} found
only a weak dependence of pulsation amplitude to He abundance in the range of
$Y_{\rm{S}} = 0.26, 0.27, 0.28$ (note that the predicted enhancement,
$\Delta Y_{\rm{S}}$, due to rotation can reach up to 0.05, cf.
Fig.\,\ref{fig:RotEnhancement}).
Furthermore, pulsation amplitudes might be expected to decrease with increased
$Y_{\rm{S}}$ due to lowered opacity in the partial ionization zone (G. Bono,
priv.
comm.).
In conclusion, pulsation amplitudes probably depend (in one way or another) on
$Y_{\rm{S}}$, which is affected by rotation. Hence, the observed spread in
observed photometric and velocimetric amplitudes \citep[e.g.][]{2000ApJ...529..293B} may
partly be due to variations in rotation rate. However, the present static models
cannot by themselves predict the behavior of pulsations.

\subsubsection{CNO elements}
In the intermediate-mass stars considered here, H-burning is dominated by
the CNO cycle. Besides producing helium from hydrogen, the CNO cycle
gradually transforms C and O into $^{14}$N, since this element has the
slowest rate of destruction \citep{1985ESOC...21..187M}. Hence, the ratios
$\rm{[N/C]}$ and $\rm{[N/O]}$ are expected to be modified by rotation. 
We show the behavior of $\Delta\,\rm{[N/C]}$ and $\Delta\,\rm{[N/O]}$ 
as a function of rotation rate in the two lower panels of Fig.
\ref{fig:RotEnhancement}.

The TAMS enhancement is predicted to be much more noticeable for CNO elements
than for helium, due to the fact that very early during the core H-burning
phase, strong gradients of CNO elements appear between the convective core and
the envelope triggering fast diffusion timescales.
Clearly, the MS enhancement of nitrogen depends primarily on rotation, as seen
by the upward facing triangles in Figs.\,\ref{fig:NCenhMass} and
\ref{fig:NOenhMass} that are grayscaled for rotation rates (faster is darker).
For fast rotating models of sufficiently high mass ($> 7\,M_\odot$) the
rotational mixing during the MS leads to stronger abundance enhancement
than dredge-up in non-rotating models of the same mass. It furthermore 
appears that the contribution of dredge-up to enhanced abundances is rather
insensitive to rotation, though the MS enhancement is very sensitive to
rotation. This is an interesting contrast with helium abundance enhancement,
pointing out the large difference between the He and CNO abundance gradients
between the convective core and the radiative envelope. In summary,
the modification of CNO abundances due to rotational mixing during the
MS and due to dredge up are of similar orders, with the rotational effect
dominating for the highest-mass models. In contrast, helium enhancement occurs
primarily after dredge-up.

We mention here that puzzling CNO abundance predictions made by rotating
Geneva models of blue supergiant stars were previously pointed out by
\citet{2013MNRAS.433.1246S}.
Although a detailed comparison with observed abundance enhancement will be
presented in a publication to follow soon, we can already mention that the
enhancement of CNO elements predicted by the present models appears
systematically too high by up to $0.2$\,dex. However, measuring
CNO abundances in the extended atmospheres of Cepheids is
sufficiently complex to leave room for significant systematic errors associated
with the measurement.

\subsection{Tracing Cepheid evolution via period changes}\label{sec:Pchange}
Cepheids are among the rare objects whose evolution can be observed on human
timescales thanks to variations in period that are interpreted as being due to
contraction (period becomes shorter) or expansion (longer) as the Cepheid
evolves along the IS crossing. The underlying idea is
that in a radially pulsating star the pulsation period, $P$, is intimately
linked with the mean density in solar units, $\bar{\rho}$, via the relation
\begin{equation}
P \cdot \sqrt{\bar{\rho}} = const \equiv Q\,.
\label{eq:Prho}
\end{equation}
This relation was first investigated in the context of stellar pulsations
by \citet{1879Ritter}. Equation\,\ref{eq:Prho} was first applied to
investigate the rate of progress of stellar evolution
by \citet{1918Obs....41..379E,1919Obs....42..338E}, demonstrating that
the main energy source of stars could not be contraction. Thus, rates of period
change have provided and continue to provide crucial tests of stellar evolution.

Estimates for predicted rates of period change can be obtained from 
evolutionary tracks such as the present ones, since
variations in mean density are time resolved during the IS crossings. 
Starting from the time derivative of the period-density relation (Eq.
\ref{eq:Prho}), we obtain
\begin{equation}
\frac{\mathrm{d}P}{\mathrm{d}t} = \frac{\mathrm{d}}{\mathrm{d}t}\left( Q\cdot \sqrt{\frac{1}{\bar{\rho}}} \right)
= \frac{\mathrm{d}Q}{\mathrm{d}P} \frac{\mathrm{d}P}{\mathrm{d}t} \sqrt{\frac{1}{\bar{\rho}}} -
\frac{Q}{2} \sqrt{\frac{1}{\bar{\rho^3}}} \frac{\mathrm{d}\bar{\rho}}{\mathrm{d}t}\, \rm{,\ and\ thus}
\label{eq:dPdt1}
\end{equation}
\begin{equation}
\frac{\mathrm{d}P}{\mathrm{d}t} = - \frac{1}{\left( 1 -
\frac{\mathrm{d}Q}{\mathrm{d}P}\sqrt{\frac{1}{\bar{\rho}}} \right)} \cdot \frac{Q}{2}
\sqrt{\frac{1}{\bar{\rho^3}}} \frac{\mathrm{d}\bar{\rho}}{\mathrm{d}t}\, .
\label{eq:dPdt2}
\end{equation}
Adopting the period-dependence of Q as investigated by
\citet{1998ApJ...498..360S}, i.e., 
\begin{equation}
Q = 3.47 \times 10^{-2} + 5.2 \times 10^{-3} \log{P} + 2.8
\times 10^{-3} [\log{P}]^{2}\, ,
\label{eq:pulsationconstant}
\end{equation}
with $0.035 \leq Q \leq 0.050$, we can estimate the importance of the
first term in Eq.\,\ref{eq:dPdt2} using the average densities predicted by the
models. We find this term to be close to unity and thus negligible for a first
order estimate. A more detailed investigation would, however, involve a
consistent determination of pulsation periods for these models. With
$\frac{\mathrm{d}x/\mathrm{d}t}{x} = \frac{\mathrm{d}\ln{x}}{\mathrm{d}t}$ 
and dividing by $P$, we obtain a simple way to predict rates of period change to
first order from the evolutionary tracks via 
\begin{equation}
\frac{\dot{P}}{P} = \frac{\mathrm{d}\log{P}}{\mathrm{d}t} \approx -
\frac{1}{2}\frac{\mathrm{d}\log{\bar{\rho}}}{\mathrm{d}t} \, .
\label{eq:dPdt}
\end{equation}

\begin{figure}
\includegraphics{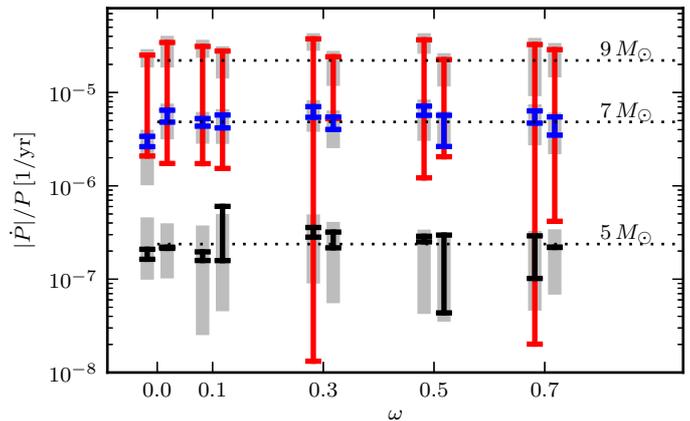}
\caption{Absolute values of predicted rates of period change for several
Cepheid models are shown against initial rotation rate $\omega$. 
Predictions for the \citet{2000ApJ...529..293B} IS are drawn as black, blue, and
red vertical lines for the $5,\,7\,$ and $9\,M_\odot$ models, respectively.
Gray shaded bars represent the predictions assuming the
\citet{2003A&A...404..423T} IS.
Second crossing Cepheids are offset to slightly
lower $\omega$, third crossing Cepheids to higher $\omega$ for
visibility.
Period decreases along the second crossing, and increases along the third.
Per-mass averages are shown as annotated
horizontal dotted lines.
}
\label{fig:dlogP/dt vs omega} 
\end{figure}
Figure\,\ref{fig:dlogP/dt vs omega} shows the range of
$\vert\dot{P}\vert/P$ estimated using Eq.\,\ref{eq:dPdt} as a function of
$\omega$ for second and third crossings. The absolute value is plotted here in
order to have the direct comparison of the rates of period change for both
crossings. 

As expected from Cepheid lifetimes, higher mass stars are predicted to show
faster rates of period change. No obvious dependence on $\omega$
emerges, although the values predicted for $\dot{P}/P$ do exhibit some
fluctuations. The only slight tendency seen is that $7\,M_\odot$
models tend to show increasing in $\dot{\vert P\vert}/P$ with increasing
$\omega$ on the second crossing, whereas the opposite is true on the third
crossing. 
Furthermore, the ranges of $\dot{P}/P$ covered in a
given crossing depend quite strongly on the IS boundaries adopted.

Period changes can be measured observationally using O-C
diagrams, and are typically given in terms of the quantity $\dot{P}/P$.  
In the Galaxy, observational rates of period change have been determined for
nearly 200 Cepheids \citep{2006PASP..118..410T}, providing good statistics and
long baselines that make this an exciting tool for comparisons with stellar models.
Comparing the rates of period change predicted by the present models with the
observed rates \citet[Fig.\,3]{2006PASP..118..410T} yields very good agreement.

Several effects not considered here are likely to impact the predicted rates of
period change, including for instance the distribution of helium inside the
envelope, or the dependence of the neglected first term in Eq.\,\ref{eq:dPdt2} 
on density (which depends on $\omega$).
However, consistent determinations of the IS location and pulsation periods 
are out of scope for this paper. These and a detailed comparison with observed
rates of period change are therefore postponed to a future publication.

\subsection{Intrinsic dispersion of the Cepheid PLR}
As was shown in the previous sections, rotation together with the crossing
number lifts the uniqueness of the mass-luminosity relationship for Cepheids and
affects their densities. One of the cornerstones of the universal distance scale
is the empirically-calibrated period-luminosity-relation 
\citep[PLR,][]{1908AnHar..60...87L,1912HarCi.173....1L}
that attributes luminosities (or absolute magnitudes) to values of the pulsation
period alone
\citep[e.g.][]{2007AJ....133.1810B,2012ApJ...758...24F}, or to period and color
\citep[e.g.][leading to Period-luminosity-color
relations]{2003A&A...404..423T,2004A&A...424...43S}. 
PLR calibrations are performed on real stellar populations (or at least samples
of stars), whose constituents formed following an $\omega$-distribution. 
The prediction that a Cepheid's luminosity depends on $\omega$ leads to the
question how this may affect the PLR.   

\begin{figure}
\centering
\includegraphics{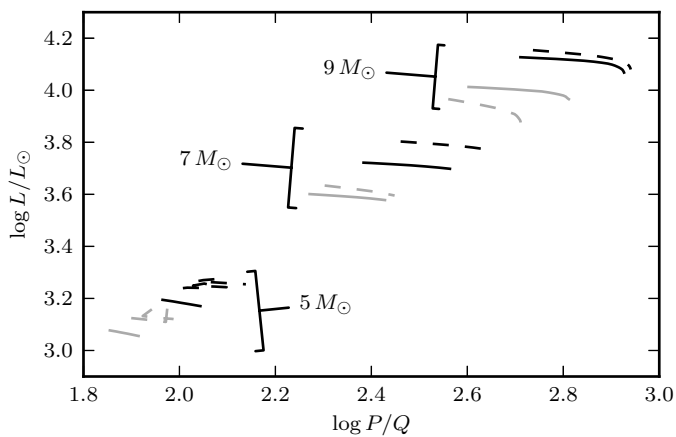}
\caption{Cepheid luminosity versus $\log{P/Q}$
for $5,\,7$ and $9\,M_\odot$ models in the second (solid lines) and third
(dashed lines) crossings. Rotating models with $\omega=0.5$ are shown in black,
non-rotating models in gray.
At a given value of $\log{P/Q}$, $\log{L}$ can easily vary by
more than 0.1 dex (25\%), depending on $\omega$ as well as the crossing number.
Hence, rotation leads to an intrinsic dispersion of the
period-luminosity-relationship.}
\label{fig:LumiDensity}
\end{figure}

Fig.\,\ref{fig:LumiDensity} shows the relation between the (logarithmic)
luminosity and the logarithm of the inverse average density
($1/\sqrt{\bar{\rho}} \equiv P/Q$, cf.
Eq.\,\ref{eq:Prho}) for the $5,\,7,$ and $9\,M_\odot$ models with ($\omega=0.5$)
and without rotation.
Second and third crossings are distinguished by line style (solid/dashed).
It is clear from the figure that a range of luminosities is
predicted for different $\omega$ values at a given $\log{P/Q}$. 
Models on different crossings can have the same effect.
Hence, the present models predict an intrinsic scatter in luminosity at
fixed $\log{P/Q}$ that is due to luminosity differences among crossings
and initial rotation rates, enabled by the finite width (in temperature) of the
IS. Naturally, if the boundaries of the IS should depend on rotation, then this may 
also affect the intrinsic dispersion of the Cepheid PLR. 
Note that color or effective temperature may likely be used to
distinguish between the various luminosities associated with a given value of
$\log{P/Q}$ \citep[e.g.][]{2003A&A...404..423T}.
 
From Fig.\,\ref{fig:LumiDensity} it appears that the maximal difference in
$\log{L}$ (at fixed $\log{P/Q}$) due to rotation can be larger than the one
due to confused crossing numbers. However, it should be remembered that luminosity is
relatively stable over a range of $\omega$ for Cepheids of a given mass, cf.
Figs.\,\ref{fig:observables_2C} and\,\ref{fig:observables_3C}. Hence, the
dispersion due to rotation alone is probably not very large, and the differences
due to the increased luminosity on the third crossing (relative to the second
crossing and at fixed $\log{P/Q}$) should be the dominant source of
dispersion in the PLR.

\section{Discussion}
	
\label{sec:Discussion}
\subsection{Rotation vs. overshoot and the mass discrepancy}
In Sec.\,\ref{sec:CepMass} we investigate Cepheid masses inferred using rotating
models and find that neglecting the effect of rotation and crossing number on
luminosity can lead to errors in the order of magnitude of the mass discrepancy.
Furthermore, the `average' M-L relationship of rotating models, i.e., neglecting
the crossing number and using the average rotation rate $\omega=0.5$, is
consistent with an M-L relation assuming significantly stronger overshooting.
This degeneracy can be easily explained, since both rotation and overshoot
increase core size and therefore produce higher luminosities for lower masses.

However, besides increased luminosity, core-overshooting makes few other
predictions, acting essentially like a fudge factor that is increased until evolutionary
masses match pulsational ones. On the other hand, rotation is observed in most
stars, and especially so in the progenitors of Cepheids. Furthermore, rotation
makes several \emph{testable} predictions of observable quantities. Accordingly,
a detailed comparison of the predictions presented here with observational
quantities is currently in preparation. 

Another possibilty of diminishing the mass discrepancy, i.e., lowering the
evolutionary mass at fixed luminosity, is to include pulsation-enhanced
mass-loss \citep[e.g.][]{2008ApJ...684..569N,2011A&A...529L...9N}. This mechanism
is not currently included in our models, and its total predicted mass loss
depends strongly on Cepheid lifetimes, i.e., on the position and width of the IS.
Observational constraints such as those by \citet{2012ApJ...744...53M} are
crucial to further investigate this issue.

\subsection{Dependence on IS boundary definition}
Clearly, adopting a given set of IS boundaries impacts the
predictions obtained from our static evolutionary models, yielding
IS boundary-dependent predictions for the range of luminosities, surface
gravities, equatorial velocities, etc. This problem, however, is common in
Cepheid research, complicated by model assumptions in the case of theoretical
boundaries and by the correction for reddening and extinction in observational
studies. 
In order to be sensitive to such systematic differences for the predicted
values, two different IS boundary definitions were employed,  
one theoretically \citep{2000ApJ...529..293B} and one empirically-derived
\citep{2003A&A...404..423T}. 

The predictions obtained from the two different IS boundary definitions do
generally vary quite a bit, although not by an order of magnitude. Generally
speaking, the \citet{2003A&A...404..423T} IS is bluer, i.e., hotter, and
therefore predicts more luminous Cepheids with higher surface gravity and
surface rotation. As a result, the `average' M-L relation for the
\citet{2003A&A...404..423T} IS predicts even lower mass at a given
luminosity than the \citep{2000ApJ...529..293B} IS, although the difference is
generally $<1\%$.
As seen in Fig.\,\ref{fig:dlogP/dt vs omega}, predicted rates
of period change may also depend on the IS boundaries adopted. Other
predicted quantities such as surface abundance enrichment, or Cepheid ages do
not significantly depend on  the IS definition. 

For a given position in the HRD, rotation lowers stellar mean density
hydrostatically, decreasing the temperature gradient. Thus, rotation is likely
to affect the location of the IS boundaries in the HRD, since pulsational
instability depends on the location of the partial He-ionization zone inside a
star's envelope as well as the total mass of helium in that region.
An $\omega$-dependence of IS boundaries would require to replace the canonical
notion of a single sharply-defined IS with a transition zone spanned by
instability trips corresponding to different $\omega$, with implications for the
question of  purity of the IS.
Such questions shall be addressed in a future study focusing on the pulsational
instability of the present models.

\section{Conclusions}\label{sec:Conclusions}
This paper presents the first detailed investigation of the effect of rotation
on evolutionary models of classical Cepheids. The study is based on 
the latest state-of-the-art rotating Geneva models (cf. paper\,I)
that incorporate a homogeneous and self-consistent treatment of rotation over
the entire evolutionary cycle for a wide range of stellar masses. 
A dense grid of evolutionary tracks of different rotation rates 
for intermediate-mass stars is available from paper\,II and enables a detailed
investigation not only as a function of initial rotation rate, $\omega$, but
also as a function of time during the IS crossings. 

Qualitative predictions are made as a function of initial rotation rate for an
array of observable quantities, such as surface gravities, surface abundance
enhancement, surface velocities, radii, and rates of period change. These will
be quantified and compared to observational constraints in a subsequent
publication. 

The key results of this investigation are:
\begin{enumerate}
  \item M-L relations depend on $\omega$. This is true for all stars during all
  stages of evolution, although the difference is more obvious during the blue
  loop phase.
  \item For Cepheids, an M-L relation at fixed $\omega$ furthermore depends
  on crossing number. The greater $\omega$, the greater the difference between
  the crossings, since rotation broadens the blue loops.
  \item The Cepheid mass discrepancy problem vanishes when rotation and
  crossing number are taken into account, without a need for high
  core overshooting values or enhanced mass loss.
  \item Differences in initial rotation rate and crossing number between
  Cepheids of identical mass and metallicity create
  intrinsic scatter around the average values predicted. This is true
  for most parameters considered here, among them in particular the Cepheid
  period-luminosity-relation, i.e., rotation is a source of intrinsic dispersion
  for the PLR.
  \item Rotational mixing can significantly enhance surface abundances during
  the MS phase. Consequently, enriched surface abundances do not unambiguously
  distinguish Cepheids on the first crossing from ones on second or third
  crossings.
  \item Rotating models predict older Cepheids than currently assumed due to
  longer MS lifetimes.
\end{enumerate}

Arguably the two most important results are result numbers 3 and 4, i.e., the
solution of the mass discrepancy problem and the prediction of intrinsic
scatter for Cepheid observables at fixed mass and metallicity.
Since the mass discrepancy has been claimed to be solved by several other
studies in the past, one may be reluctant to immediately adopt
rotation as the best explanation.
However, result 4 may provide the smoking gun for distinguishing between
rotation and other effects such as core overshooting.
While it would be difficult to
explain why stars of identical mass and metallicity should present different
overshooting values (leading to the observed scatter in, say, radius or
luminosity), dispersion arises naturally when a dispersion in initial
velocity is considered, which is an observational fact. 

As this paper shows, the effects of rotation on Cepheid populations are highly
significant, ranging from issues related to inferred masses to the prediction of
systematic effects relevant for the distance scale. Further related research is
currently in progress.

\begin{acknowledgements}
The authors would like to acknowledge Nancy Remage Evans for her communication
regarding model-independent mass-estimates of classical Cepheids and the
referee, Giuseppe Bono, for his valuable comments. This
research has made use of NASA's Astrophysics Data System.
CG acknowledges support from the European Research Council under the European
Union's Seventh Framework Programme (FP/2007-2013) / ERC Grant Agreement n.
306901.   
\end{acknowledgements}


\bibliographystyle{aa} 
\bibliography{cepheid_model_library,Bib_StellarEvolution,Bib_DistanceScale}    

\end{document}